\documentclass{article} % For LaTeX2e
\usepackage{iclr2025_conference,times}

\usepackage{amsmath,amsfonts,bm}

\def\eqref#1{equation~\ref{#1}}
\def\1{\bm{1}}

\DeclareMathAlphabet{\mathsfit}{\encodingdefault}{\sfdefault}{m}{sl}
\SetMathAlphabet{\mathsfit}{bold}{\encodingdefault}{\sfdefault}{bx}{n}

\usepackage{hyperref}
\usepackage{url}
\usepackage{graphicx}
\usepackage{booktabs}
\usepackage{subcaption}

\newcommand{\code}{https://github.com/finngueterbock/FYP}

\usepackage{comment}

\title{Improving Local Air Quality Predictions Using Transfer Learning on Satellite Data and Graph Neural Networks}

\author{Finn Gueterbock  \\
University of Bristol, UK \\
\And
Raul Santos-Rodriguez \\
University of Bristol, UK \\
\And
Jeffrey N. Clark \\
University of Bristol, UK \\
\texttt{jeff.clark@bristol.ac.uk} \\
}

\iclrfinalcopy % Uncomment for camera-ready version, but NOT for submission.
\begin{document}

\maketitle

\begin{abstract}
Air pollution is a significant global health risk, contributing to millions of premature deaths annually. Nitrogen dioxide (NO$_2$), a harmful pollutant, disproportionately affects urban areas where monitoring networks are often sparse. We propose a novel method for predicting NO$_2$ concentrations at unmonitored locations using transfer learning with satellite and meteorological data. Leveraging the GraphSAGE framework, our approach integrates autoregression and transfer learning to enhance predictive accuracy in data-scarce regions like Bristol. Pre-trained on data from London, UK, our model achieves a 8.6\% reduction in Normalised Root Mean Squared Error (NRMSE) and a 32.6\% reduction in Gradient RMSE compared to a baseline model. This work demonstrates the potential of virtual sensors for cost-effective air quality monitoring, contributing to actionable insights for climate and health interventions.
\end{abstract}

\section{Introduction}
Air pollution is one of the leading causes of global mortality, responsible for over 8 million deaths annually \citep{lelieveld2023air}. Nitrogen dioxide (NO$_2$), primarily emitted from vehicles and industrial activities, has severe health impacts, particularly in urban environments. Despite its significance, the limited number of ground-based monitoring stations hinders the ability to measure air quality effectively, mitigate localised air quality challenges effectively, and evaluate the impact of policies \citep{who2018}.

Satellite-based data provides global coverage of air quality metrics; however, its low spatial resolution restricts its applicability for localised decision-making. To address this gap, we propose a novel model that combines satellite and meteorological data with sparse ground-based measurements to simulate high-resolution NO$_2$ readings at unmonitored locations, creating `virtual sensors'. %This approach would enable more informed decision-making for public health authorities, particularly in areas with limited resources for local air quality monitoring.

We introduce an inductive learning framework based on GraphSAGE \citep{hamilton} that incorporates temporal dependencies through autoregression and enhances performance using transfer learning. By leveraging transfer learning from cities with better monitoring networks, this work seeks to improve the accuracy and generalizability of the model, enabling it to predict air quality in regions with fewer monitoring stations. We achieve sizeable accuracy improvements in unseen locations. This approach has the potential to enable low-cost, scalable air quality monitoring, especially in resource-constrained regions, thereby supporting global efforts toward better health outcomes and climate resilience.

\section{Related Work}
Air quality prediction has been extensively studied using a variety of machine learning techniques. Traditional methods often rely on geostatistical interpolation, such as kriging \citep{kriging1990}. While effective for some spatial analyses, these methods struggle with the complex spatiotemporal relationships in air quality data. Graph Neural Networks (GNNs) offer a more flexible approach by leveraging the relationships between data points in a network \citep{xu2018powerful}. For example, \citet{muthukumar2022} used a Graph Convolutional Network (GCN) combined with time series models to predict PM\textsubscript{2.5} levels, demonstrating the potential of GNNs for air quality forecasting.

Satellite data has become a key resource in air quality research, offering global coverage of atmospheric metrics such as Aerosol Optical Depth (AOD) and NO$_2$ column density. \citet{masih2019} demonstrated that machine learning models, such as Random Forests, can predict NO$_2$ concentrations from satellite and meteorological data. Similarly, \citet{deep2021} applied deep learning to estimate daily NO$_2$ levels, achieving promising results. However, these approaches often lack the spatial resolution necessary for accurate local predictions, particularly in urban environments.

Transfer learning has shown potential for improving model generalisability, especially in data-scarce regions. \citet{ma2019improving} found that transfer learning improved air quality prediction accuracy when applied to larger temporal resolutions. \citet{accra-transfer} utilised deep transfer learning on satellite imagery to enhance air quality predictions in developing countries, demonstrating the feasibility of pre-training on well-resourced cities and fine-tuning on data-poor areas. Despite these advancements, the high spatial resolution necessary for accurate NO$_2$ predictions in urban environments remains a challenge.

This study aims to bridge this gap by using transfer learning and GNN-based models to provide more accurate, location-specific NO$_2$ predictions, addressing limitations in both spatial and temporal resolution present in previous work.

\section{Methods}
\subsection{Data}
\label{sec:data}

\begin{figure}[!htbp]
\centering
\begin{subfigure}{.5\textwidth}
  \centering
  \includegraphics[width=\linewidth]{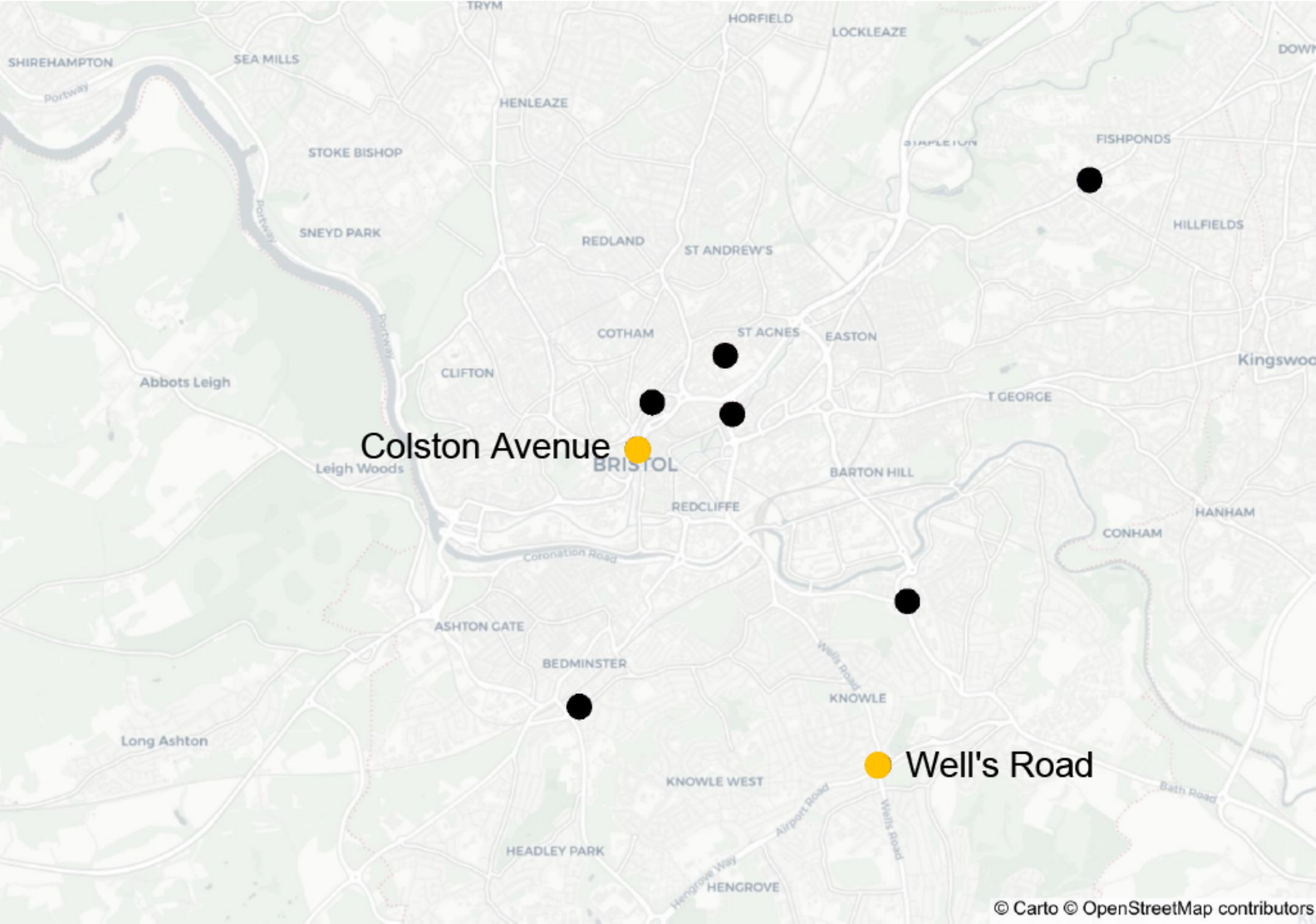}
  \caption{Bristol (n = 8)}
  \label{fig:bristol_map}
\end{subfigure}%
\begin{subfigure}{.5\textwidth}
  \centering
  \includegraphics[width=0.88\linewidth]{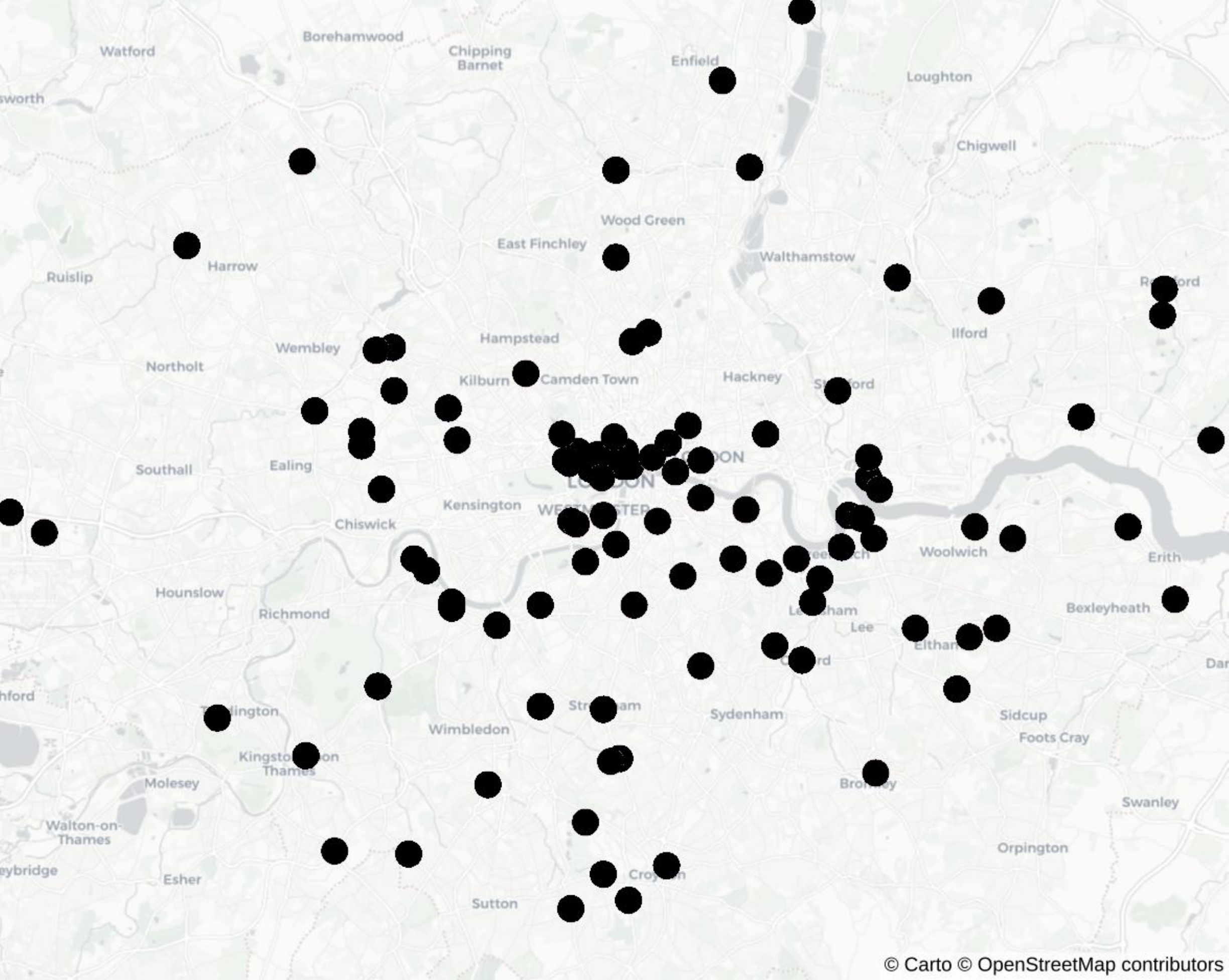}
  \caption{London (n = 112)}
  \label{fig:london_map}
\end{subfigure}
\caption{Maps showing the spatial distribution of NO$_2$ sensors locations in (a) Bristol and (b) London accessed during this study. In (a), the two Bristol sensor locations marked in orange are those for which temporal NO$_2$ predictions are presented in \autoref{fig:predictions_subfigures}.}
\label{fig:sensor_map}
\end{figure}

\paragraph{Surface NO$_2$ Measurements}
Ground-based NO$_2$ measurements for the Bristol area were obtained from the Air Quality Data Continuous dataset via the Open Data Bristol API \citep{bristoldata}, which provides hourly air quality data from 19 different locations across the city since 1993. For London, we use data from the London Air Quality Network \cite{mittal2020}, which reports hourly NO$_2$ readings from 112 locations across Greater London. From 2018 onwards, the period of time for which satellite data is available, the datasets includes 246,572 data points across 8 locations in Bristol, and 4,182,699 across 112 locations in London (\autoref{fig:sensor_map}).

\paragraph{Satellite Data}
Satellite data from the European Space Agency's Sentinel-5P (2018–present) provides daily NO$_2$ concentrations and aerosol indices at a 5.5 x 3.5 km resolution. The satellite data is treated as static between daily measurements to match its temporal resolution.

\paragraph{Additional Features}
Hourly meteorological data from the ERA5-Land dataset \cite{weather_data} including variables such as temperature, wind, humidity, and cloud cover at a 5 km resolution were included. All features are listed in Appendix~\ref{ap:dataset:data}. Proximity to A-roads or motorways was calculated using OS Open Roads \cite{osopenroads}, providing a distance-to-road feature for each sensor location. All features were standardised.

\subsection{Models}
\label{sec:models}

GraphSAGE aggregates data from a sensor's local neighborhood, improving its ability to capture geographical variations. We introduce autoregression to model temporal dependencies in NO$_2$ concentrations, as shown by the data autocorrelation (Appendix~\ref{ap:dataset:data},
\autoref{fig:autocorrelation}). This enables the model to make more accurate predictions by considering past NO$_2$ values. Transfer learning was conducted by pre-training on London data and fine-tuning on Bristol data. Performance was assessed against models trained on only Bristol data. For comparison purposes we evaluate three additional non-GNN models: XGBoost, MLP, and CNN, and performed transfer learning on the best performing of these baseline models. Appendix \ref{ap:dataset:models} details model training.

Each model takes as input satellite, meteorological, and time-based features (Appendix~\ref{ap:dataset:data}) and outputs hourly NO$_2$ predictions. We test each model on unseen locations, using RMSE (Root Mean Squared Error), NRMSE (Normalized RMSE), and Grad-RMSE (Gradient RMSE) averaged across all unseen locations to assess performance as a `virtual sensor'. Each model is tested on one location at a time, having trained on all the remaining locations.  Results are presented for sample periods of time spanning several weeks. Code for all models and plots discussed in this report are available online \footnote{\url{\code}}.

\section{Results and Discussion}

The GraphSAGE model achieved an RMSE of 17.016 $\mu g/m^3$, NRMSE of 0.526, and Grad-RMSE of 9.426 $\mu g/m^3$, demonstrating improved performance compared to the MLP, XGBoost and CNN baselines (\autoref{tab:all_models}). Transfer learning from London improved predictions for both the CNN and GraphSAGE architectures, highlighting its potential for areas with fewer monitoring stations.

\begin{table}[h]
\centering
\begin{tabular}{lccc}
    \toprule
    \textbf{Model} & \textbf{RMSE} \( \downarrow \) & \textbf{NRMSE} \( \downarrow \) & \textbf{Grad-RMSE} \( \downarrow \) \\ \hline
    MLP                  & 27.482 & 0.876 & 10.812          \\
    XGBoost              & 22.773 & 0.721 & 9.583           \\
    CNN                  & 21.133 & 0.672 & 9.741           \\
    Transferred CNN      & 18.362 & 0.583 & 9.912           \\
    GraphSAGE            & 17.016 & 0.526 & 9.426           \\
    Transferred GraphSAGE & \textbf{15.623} & \textbf{0.481}  & \textbf{6.354}  \\
    \bottomrule
\end{tabular}
\caption{Performance of all models, averaged across Bristol sensor locations (n = 8).}
\label{tab:all_models}
\end{table}

The transferred GraphSAGE model achieved the best performance across all models (\autoref{tab:all_models}), with transfer learning reducing error metrics by between 8.2\% and 32.6\% (Appendix~\ref{ap:model:perf}, \autoref{tab:graph_comparisons}). The model achieved an RMSE of 15.623 $\mu g/m^3$ and an NRMSE of 0.481, both of which can be considered acceptable within the context of urban NO$_2$ forecasting, especially given the complexities involved with applying transfer learning to a new geographical area. Recent comparable studies utilising similar graph-based and hybrid methodologies typically report RMSE values in the range of 10–20 $\mu g/m^3$, with NRMSE frequently between 30\% and 50\% \citep{wang2023hybrid, qi2023graph}. Although the errors reported here are somewhat higher than those achieved by specialised, locally trained models, the demonstrated accuracy remains promising for practical application in urban environments with sparse monitoring infrastructure. In particular, there is potential for creating virtual sensors in data-scarce regions, improving predictions by capturing both spatial and temporal dependencies. The addition of other data sources, such as terrain and land-use, could be incorporated to further improve performance.

\begin{figure}[h]
\centering
\begin{subfigure}[t]{\textwidth}
\centering
\includegraphics[width=\textwidth]{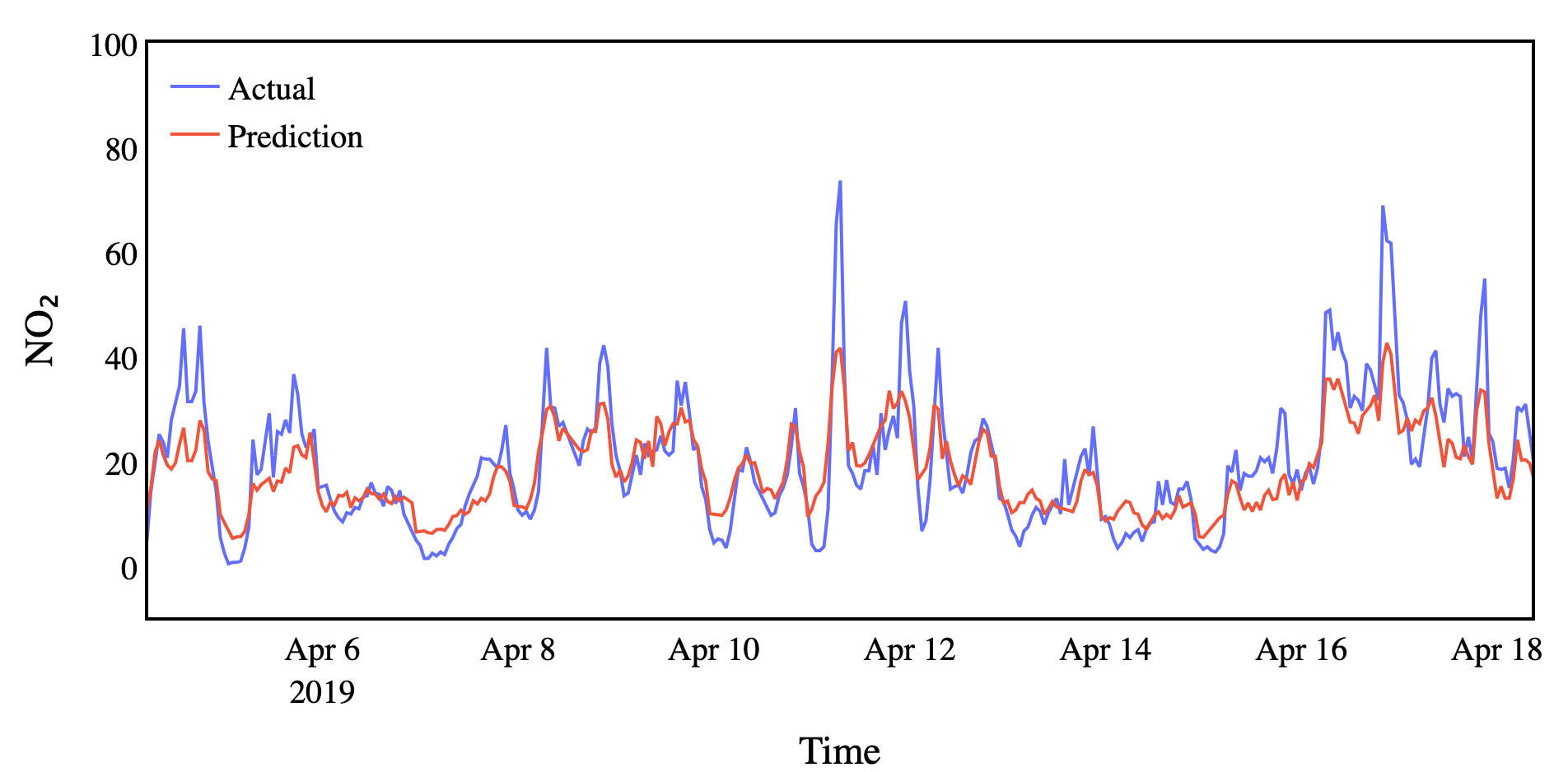}
\caption{NO$_2$ predictions for Well's Road, Bristol.}
\label{fig:wells_road_predictions}
\end{subfigure}
~
\begin{subfigure}[t]{\textwidth}
\centering
\includegraphics[width=\textwidth]{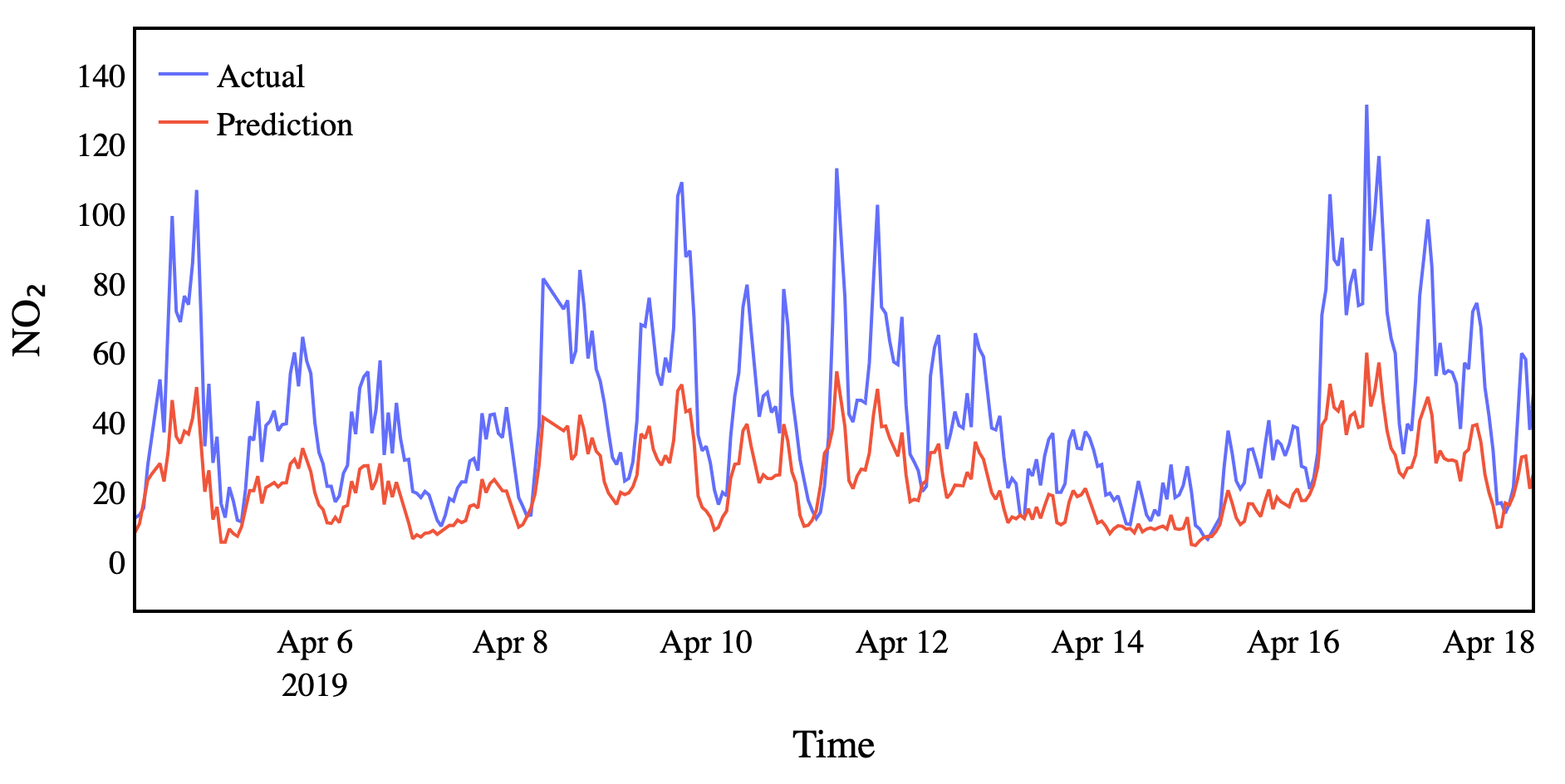}
\caption{NO$_2$ predictions for Colston Avenue, Bristol.}
\label{fig:colston_predictions}
\end{subfigure}
\caption{Actual and predicted NO$_2$ values from the transferred GraphSAGE model for two locations in Bristol: (a) Well's Road  - a location with typically low NO$_2$ values, and (b) Colston Avenue - a location with typically high NO$_2$ values.}
\label{fig:predictions_subfigures}
\end{figure}

Samples of predicted versus actual NO$_2$ values for two locations (\autoref{fig:predictions_subfigures}), demonstrate temporal prediction quality over the period of several weeks, as well as the tendency to under-predict higher NO$_2$ values. This effect is made especially clear in \autoref{fig:colston_predictions}, where the predicted values seem to follow the true values accurately, but are scaled down by some factor, leading to an overall bias. Future work should aim to address this bias, in addition to evaluating on additional metrics to measure the systematic error, both of which will be key to ensuring that the model predictions can be used to derive reliable quantitative measures of air pollutants at various scales.

%location 1 - Well's Road
%location 2- Colston Avenue

While the results are promising, challenges remain, particularly around computational efficiency. The need to reconstruct the graph for each timestep imposes significant costs, which future work should address through optimised architectures. Given the strong temporal dependence observed in NO$_2$  (Appendix~\ref{ap:dataset:data},
\autoref{fig:autocorrelation}), benchmarking against a time-series model in future work may prove valuable. We acknowledge that nearby locations may exhibit correlated air quality, which could lead to optimistic performance estimates under the employed leave-one-out evaluation. Future work should explore alternative evaluation schemes. Additionally, validation in diverse geographical regions should be carried out to assess the model's generalisability.

\section{Impact and Implications}
The proposed approach addresses critical gaps in urban air quality monitoring by enabling high-resolution NO$_2$ predictions in areas with sparse monitoring networks. This work has direct implications for public health and climate policy, providing low-cost tools for assessing air quality in resource-limited settings. By integrating satellite data with machine learning, this method supports global efforts to achieve SDG 3 (Good Health and Well-being) and SDG 13 (Climate Action)~\citep{lee2016transforming}.

Future deployments could extend to developing nations, where the lack of monitoring infrastructure exacerbates air quality challenges. There is potential to scale the approach beyond the city level to larger geographical regions. Furthermore, the approach could be applied for other pollutants for which satellite datasets are available such as methane or sulfur dioxide, enabling emissions detection and broadening the work's impact on climate resilience and urban planning. Collaborations with local governments and environmental agencies will be essential to ensure practical application and policy integration. 

\section{Conclusion}
This study demonstrates the potential of leveraging GNNs and transfer learning to address challenges in localised air quality prediction. By integrating satellite data, meteorological features, and ground-based measurements, we developed a GraphSAGE-based model capable of accurately predicting NO$_2$ concentrations at unmonitored locations in Bristol. Pre-training on data from London and fine-tuning on Bristol significantly improved model performance, achieving an 8.6\% reduction in NRMSE and a 32.6\% reduction in Gradient RMSE compared to the baseline.

Our findings highlight the feasibility of deploying virtual sensors in resource-limited settings, contributing to scalable, low-cost air quality monitoring solutions. This approach provides actionable insights for public health and urban planning, especially in cities with sparse monitoring networks.

Future work will focus on optimising the model’s computational efficiency and expanding validation across diverse geographical regions. Additionally, extending this framework to other pollutants and integrating real-time monitoring data could further enhance its utility. By addressing these challenges, this methodology has the potential to support global efforts in mitigating air pollution and combating climate change.

\begin{comment}
\subsubsection*{Author Contributions}
If you'd like to, you may include  a section for author contributions as is done
in many journals. This is optional and at the discretion of the authors.

\subsubsection*{Acknowledgments}
Use unnumbered third level headings for the acknowledgments. All
acknowledgments, including those to funding agencies, go at the end of the paper.
\end{comment}

\subsubsection*{Acknowledgments}
JNC and RSR are funded by the Self-Learning Digital Twins for Sustainable Land Management grant [grant number EP/Y00597X/1] and the UKRI Turing AI Fellowship [grant number EP/V024817/1]. 

\bibliography{iclr2025_conference}

\begin{thebibliography}{21}
\providecommand{\natexlab}[1]{#1}
\providecommand{\url}[1]{\texttt{#1}}
\expandafter\ifx\csname urlstyle\endcsname\relax
  \providecommand{\doi}[1]{doi: #1}\else
  \providecommand{\doi}{doi: \begingroup \urlstyle{rm}\Url}\fi

\bibitem[{Bristol City Council}(2022)]{bristoldata}
{Bristol City Council}.
\newblock Air quality data continuous, 2022.
\newblock URL \url{https://opendata.bristol.gov.uk/explore/dataset/air-quality-data-continuous/information/?disjunctive.location}.
\newblock [Last accessed 28/10/22].

\bibitem[Chatzidiakou et~al.(2019)Chatzidiakou, Krause, Popoola, Di~Antonio, Kellaway, Han, Squires, Wang, Zhang, Wang, et~al.]{pams}
Lia Chatzidiakou, Anika Krause, Olalekan~AM Popoola, Andrea Di~Antonio, Mike Kellaway, Yiqun Han, Freya~A Squires, Teng Wang, Hanbin Zhang, Qi~Wang, et~al.
\newblock Characterising low-cost sensors in highly portable platforms to quantify personal exposure in diverse environments.
\newblock \emph{Atmospheric measurement techniques}, 12\penalty0 (8):\penalty0 4643--4657, 2019.

\bibitem[Chen \& Guestrin(2016)Chen and Guestrin]{xgboost}
Tianqi Chen and Carlos Guestrin.
\newblock Xgboost: A scalable tree boosting system.
\newblock In \emph{Proceedings of the 22nd acm sigkdd international conference on knowledge discovery and data mining}, pp.\  785--794, 2016.

\bibitem[Cressie(1990)]{kriging1990}
Noel Cressie.
\newblock The origins of kriging.
\newblock \emph{Mathematical geology}, 22\penalty0 (3):\penalty0 239--252, 1990.

\bibitem[Data61(2018)]{StellarGraph}
CSIRO's Data61.
\newblock Stellargraph machine learning library.
\newblock \url{https://github.com/stellargraph/stellargraph}, 2018.

\bibitem[Ghahremanloo et~al.(2021)Ghahremanloo, Lops, Choi, and Yeganeh]{deep2021}
Masoud Ghahremanloo, Yannic Lops, Yunsoo Choi, and Bijan Yeganeh.
\newblock Deep learning estimation of daily ground-level no2 concentrations from remote sensing data.
\newblock \emph{Journal of Geophysical Research: Atmospheres}, 126\penalty0 (21):\penalty0 e2021JD034925, 2021.

\bibitem[Hamilton et~al.(2017)Hamilton, Ying, and Leskovec]{hamilton}
William~L. Hamilton, Rex Ying, and Jure Leskovec.
\newblock Inductive representation learning on large graphs.
\newblock In \emph{Proceedings of the 31st International Conference on Neural Information Processing Systems}, NIPS'17, pp.\  1025–1035, 2017.
\newblock ISBN 9781510860964.

\bibitem[Lee et~al.(2016)Lee, Kjaerulf, Turner, Cohen, Donnelly, Muggah, Davis, Realini, Kieselbach, MacGregor, et~al.]{lee2016transforming}
Bandy~X Lee, Finn Kjaerulf, Shannon Turner, Larry Cohen, Peter~D Donnelly, Robert Muggah, Rachel Davis, Anna Realini, Berit Kieselbach, Lori~Snyder MacGregor, et~al.
\newblock Transforming our world: implementing the 2030 agenda through sustainable development goal indicators.
\newblock \emph{Journal of public health policy}, 37:\penalty0 13--31, 2016.

\bibitem[Lelieveld et~al.(2023)Lelieveld, Haines, Burnett, Tonne, Klingm{\"u}ller, M{\"u}nzel, and Pozzer]{lelieveld2023air}
Jos Lelieveld, Andy Haines, Richard Burnett, Cathryn Tonne, Klaus Klingm{\"u}ller, Thomas M{\"u}nzel, and Andrea Pozzer.
\newblock Air pollution deaths attributable to fossil fuels: observational and modelling study.
\newblock \emph{bmj}, 383, 2023.

\bibitem[Ma et~al.(2019)Ma, Cheng, Lin, Tan, and Zhang]{ma2019improving}
Jun Ma, Jack~CP Cheng, Changqing Lin, Yi~Tan, and Jingcheng Zhang.
\newblock Improving air quality prediction accuracy at larger temporal resolutions using deep learning and transfer learning techniques.
\newblock \emph{Atmospheric Environment}, 214:\penalty0 116885, 2019.

\bibitem[Masih(2019)]{masih2019}
Adven Masih.
\newblock Application of random forest algorithm to predict the atmospheric concentration of no2.
\newblock In \emph{2019 Ural Symposium on Biomedical Engineering, Radioelectronics and Information Technology (USBEREIT)}, pp.\  252--255. IEEE, 2019.

\bibitem[Mittal(2020)]{mittal2020}
Louise Mittal.
\newblock London air quality network summary report, 2020.
\newblock URL \url{https://londonair.org.uk/london/reports/2020_LAQN_Report.pdf}.
\newblock [Last accessed 28/11/22].

\bibitem[Mu{\~n}oz-Sabater et~al.(2021)Mu{\~n}oz-Sabater, Dutra, Agust{\'\i}-Panareda, Albergel, Arduini, Balsamo, Boussetta, Choulga, Harrigan, Hersbach, et~al.]{weather_data}
Joaqu{\'\i}n Mu{\~n}oz-Sabater, Emanuel Dutra, Anna Agust{\'\i}-Panareda, Cl{\'e}ment Albergel, Gabriele Arduini, Gianpaolo Balsamo, Souhail Boussetta, Margarita Choulga, Shaun Harrigan, Hans Hersbach, et~al.
\newblock Era5-land: A state-of-the-art global reanalysis dataset for land applications.
\newblock \emph{Earth System Science Data}, 13\penalty0 (9):\penalty0 4349--4383, 2021.

\bibitem[Muthukumar et~al.(2022)Muthukumar, Cocom, Nagrecha, Comer, Burga, Taub, Calvert, Holm, and Pourhomayoun]{muthukumar2022}
Pratyush Muthukumar, Emmanuel Cocom, Kabir Nagrecha, Dawn Comer, Irene Burga, Jeremy Taub, Chisato~Fukuda Calvert, Jeanne Holm, and Mohammad Pourhomayoun.
\newblock Predicting pm2. 5 atmospheric air pollution using deep learning with meteorological data and ground-based observations and remote-sensing satellite big data.
\newblock \emph{Air Quality, Atmosphere \& Health}, 15\penalty0 (7):\penalty0 1221--1234, 2022.

\bibitem[{Ordnance Survey}(2022)]{osopenroads}
{Ordnance Survey}.
\newblock Os open roads.
\newblock \url{https://www.ordnancesurvey.co.uk/business-government/products/open-roads.html}, 2022.

\bibitem[Qi et~al.(2023)Qi, Liu, Zhang, Huang, and Zhao]{qi2023graph}
Junyu Qi, Bin Liu, Lei Zhang, Jie Huang, and Honglei Zhao.
\newblock Graph neural network-based spatiotemporal air quality prediction model with satellite remote sensing and meteorological data.
\newblock \emph{Environmental Pollution}, 316:\penalty0 120596, 2023.

\bibitem[Thomas \& Gunner(2023)Thomas and Gunner]{localair}
James Thomas and Sam Gunner.
\newblock Local air – mapping the local environment using e‑scooters, 2023.
\newblock URL \url{https://www.localair.uk/}.

\bibitem[Wang et~al.(2023)Wang, Li, Zhang, Liu, and Huang]{wang2023hybrid}
Yujie Wang, Shuang Li, Ke~Zhang, Hongyu Liu, and Wei Huang.
\newblock A hybrid deep learning approach for no$_2$ concentration forecasting combining transformer and lstm.
\newblock \emph{Atmospheric Pollution Research}, 14\penalty0 (2):\penalty0 101615, 2023.

\bibitem[{World Health Organization and others}(2018)]{who2018}
{World Health Organization and others}.
\newblock Air pollution and child health: prescribing clean air: summary.
\newblock Technical report, World Health Organization, 2018.

\bibitem[Xu et~al.(2018)Xu, Hu, Leskovec, and Jegelka]{xu2018powerful}
Keyulu Xu, Weihua Hu, Jure Leskovec, and Stefanie Jegelka.
\newblock How powerful are graph neural networks?
\newblock \emph{arXiv preprint arXiv:1810.00826}, 2018.

\bibitem[Yadav et~al.(2022)Yadav, Sorek-Hamer, Von~Pohle, Asanjan, Sahasrabhojanee, Suel, Arku, Lingenfelter, Brauer, Ezzati, et~al.]{accra-transfer}
Nishant Yadav, Meytar Sorek-Hamer, Michael Von~Pohle, Ata~Akbari Asanjan, Adwait Sahasrabhojanee, Esra Suel, Raphael Arku, Violet Lingenfelter, Michael Brauer, Majid Ezzati, et~al.
\newblock Deep transfer learning on satellite imagery improves air quality estimates in developing nations.
\newblock \emph{arXiv preprint arXiv:2202.08890}, 2022.

\end{thebibliography}
\bibliographystyle{iclr2025_conference}

\appendix
\section{Appendix}

\subsection{Data} \label{ap:dataset:data}

The following features were used to predict local NO$_2$ measurements:

\textbf{Satellite Features:} Tropospheric NO$_2$ column number density, absorbing aerosol index; \\
\textbf{Meteorological Features:} Wind speed, wind gust speed, wind direction, vapour pressure deficit, temperature, surface pressure, relative humidity, dew point, cloud cover percentage;\\
\textbf{Time-based Features:} Day of the week, week of the year and time of day. 

\begin{figure}[!htbp]
    \centering
    \includegraphics[width=.6\textwidth]{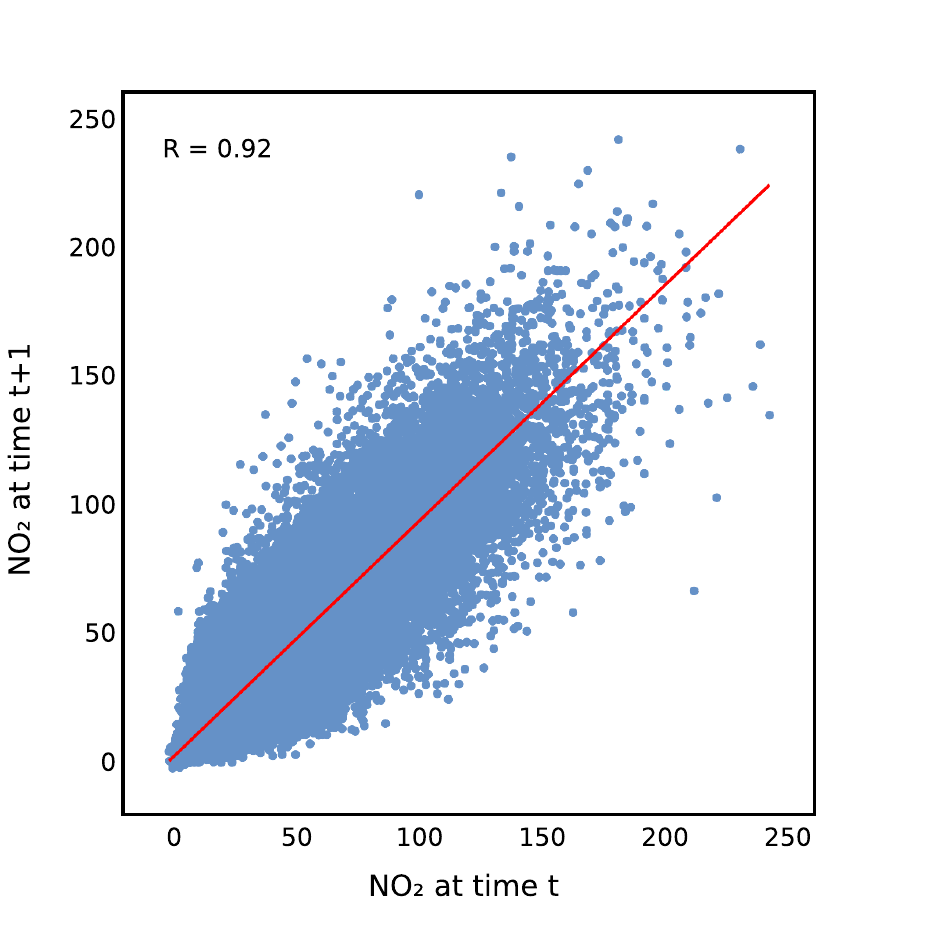}
    \caption{NO$_2$ values at time $t$ vs NO$_2$ values at time $t+1$.}
    \label{fig:autocorrelation}
\end{figure}

\subsection{Model training} \label{ap:dataset:models}
\paragraph{GraphSAGE}
Since the GraphSAGE algorithm \citep{hamilton} works by sampling from a node’s local neighbourhood of connected nodes, feeding in the data all at once in this way does not allow the model to see other timesteps in the future or the past, as the graphs at each timestep are not connected to each other.

In order to address this issue of not having a continuous representation of time in the graph, we include the predicted NO$_2$ value from a node's previous timestep as an input for the current prediction in a process called autoregression. Autoregression is a technique used to model time series data, where each data point is predicted based on the previous predicted values. During training, we include the previous timestep's actual NO$_2$ value at each node as a feature for that node. When nodes at a certain timestep are missing, the last recorded value for that time of day is used in its place. This should not greatly affect our results due to the high correlation the NO$_2$ values have with time. 
This enables the model to learn how to aggregate the satellite data, meteorological data and previous NO$_2$ values for each node and its neighbours. The use of autoregression is particularly effective for modeling time series data that exhibit a high degree of autocorrelation, as is the case with the NO$_2$ concentration data. Autocorrelation is a measure of the degree to which a data point is correlated with its preceding data points, and can be visualised using an autocorrelation plot (\autoref{fig:autocorrelation}). As can be observed from the plot, there is a significant degree of correlation between the NO$_2$ concentration at each timestep and the NO$_2$ concentration at the next timestep.

To predict on an unseen node, we must first initialise the node with a value for the NO$_2$ at the previous timestep. During development of the model, this was achieved by including the actual NO$_2$ value for the first timestep, however in reality this initial sample could be provided by an air quality sampling scheme such as the Breathe London scheme, which involves children wearing portable air quality monitors on their backpacks \cite{pams}, or LocalAir, an e-scooter based air quality monitoring scheme in Bristol \cite{localair}. These methods of sampling are designed to be cheap and versatile in their applications so could be used in countries lacking infrastructure to provide a baseline NO$_2$ reading for a new location. If no such sample is available, it would also be possible to simply provide an estimate for the NO$_2$ concentration at a particular location, as the model only has access to a single previous value, forcing it to focus on the change in NO$_2$ at each timestep.

We compare the performance of four different node aggregator functions, mean, max pooling, mean pooling and attentional aggregator, as defined in the StellarGraph python library \cite{StellarGraph}, and chose mean pooling. To improve the model's ability to generalise to unseen data, we include a dropout layer with a drop out rate of 0.5 before the activation layer.

Other parameters for the model include the number of hops away from each node to sample from, the maximum number of nodes to sample at each hop and the number of neurons to use when aggregating each node and its neighbours. Since the maximum number of hops possible from any node in the Bristol graph is 2, it was decided that we would perform 2 aggregations, sampling nodes at one hop, then two hops from each node. At each of these steps, a maximum of 3 and 5 nodes would be sampled respectively. Other parameters for the model such as the number of neurons for aggregation and the learning rate were selected by trial and error. 

\paragraph{Non-GNN baseline models}
Baseline model parameters were as follows:

\begin{itemize}
\item an XGBoost model \cite{xgboost} with 100 decision trees;
\item a multilayer perceptron model (MLP) with 2 fully connected layers, a dropout layer with a rate of 0.5, and 2 more fully connected layers;
\item a convolutional neural network (CNN) model with 2 convolutional layers, a dropout layer with a rate of 0.5, and 2 fully connected layers.
\end{itemize}

\subsection{Model performance} \label{ap:model:perf}
Table~\ref{tab:graph_comparisons} illustrates the performance improvements observed using transfer learning.

\begin{table}[!htbp]
\centering
\begin{tabular}{lccc}
    \toprule
    \textbf{Model} & \textbf{RMSE} \( \downarrow \) & \textbf{NRMSE} \( \downarrow \) & \textbf{Grad-RMSE} \( \downarrow \) \\ \hline
    GraphSAGE              & 17.016 & 0.526  & 9.426     \\
    Transferred GraphSAGE  & 15.623 & 0.481  & 6.354     \\ \hline
    Percentage Improvement  & 8.185 & 8.576  & 32.593    \\
    \bottomrule
\end{tabular}
\caption{Comparison of NO$_2$ prediction performance between GraphSAGE and Transferred GraphSAGE models.}
\label{tab:graph_comparisons}
\end{table}

\end{document}